# Critical Temperature Oscillations and Reentrant Superconductivity due to the FFLO like State in F/S/F Trilayers

J. Kehrle[1], V.I. Zdravkov[1,2], G. Obermeier[1], J. Garcia-Garcia[1], A. Ullrich[1], C. Müller[1], R. Morari[2], A.S. Sidorenko[2], S. Horn[1], L.R. Tagirov[1,3], R. Tidecks[1]

[1]*Institut für Physik, Universität Augsburg, D-86159 Augsburg, Germany*
[2]*Institute of Electronic Engineering and Nanotechnologies ASM, MD2028 Kishinev, Moldova*
[3]*Solid State Physics Department, Kazan Federal University, 420008 Kazan, Russia*

**Abstract**

Ferromagnet/Superconductor/Ferromagnet (F/S/F) trilayers, in which the establishing of a Fulde-Ferrell Larkin-Ovchinnikov (FFLO) like state leads to interference effects of the superconducting pairing wave function, form the core of the superconducting spin valve. The realization of strong critical temperature oscillations in such trilayers, as a function of the ferromagnetic layer thicknesses or, even more efficient, reentrant superconductivity, are the key condition to obtain a large spin valve effect, *i.e.* a large shift in the critical temperature. Both phenomena have been realized experimentally in the $Cu_{41}Ni_{59}$/Nb/$Cu_{41}Ni_{59}$ trilayers investigated in the present work.

**1. Introduction**

Since the classical paper of Bardeen, Cooper, and Schrieffer (BCS) [1] it is known that the superconducting state is established by electron pairs (Cooper pairs) with opposite spin and momenta. Magnetism destroys superconductivity. Ferromagnetism requires parallel arrangements of electron spins. Magnetic fields act pair breaking, because they break the time reversal symmetry [2]. Therefore, it was long time not believed that superconductivity can exist on a ferromagnetic background. Nevertheless, Fulde-Ferrell and Larkin-Ovchinnikov (FFLO) showed that superconductivity can establish in the presence of an exchange field [3,4]. In this case pairing occurs with opposite spins but a non-vanishing momentum of the Cooper pair. This so called FFLO state can only be observed for a very narrow set of parameters [5], which is extremely difficult to adjust in experiments [6-11].

Thin film structures of superconductors (S) and ferromagnets (F), however, allow to realize an FFLO like state. In this case the superconducting charge carriers penetrate into the ferromagnetic material, acquiring a non-zero momentum due to the different pairing mechanism, caused by the exchange field, in the F-material. This is different to the widely studied proximity effect at an S/N interface, and it is the reason that the paring wave function in the F material does not only decay exponentially into the non-superconducting material as in a nonmagnetic N metal, but oscillates in addition [12-16]. Thus, in S/F layers an interference of the superconducting pairing wave function can happen in analogy to the case of light in a Fabry-Pérot interferometer [17]. These interference effects lead to new physical observations, such as an oscillation of the critical temperature, $T_c$, if changing the thickness of the ferromagnetic layer [18,19]. Even a total extinction of superconductivity in a layered structure of superconducting and ferromagnetic materials may happen. This phenomenon was theoretically predicted [20] already a long time ago before it could be realized experimentally. Only recently, distinct oscillations of the critical temperature and a clear observation of an extinction and recovery of the superconducting state (denoted as reentrant superconducting behavior), could be demonstrated in S/F bilayers [21,22] using Nb as a superconducting material (S) and the $Cu_{49}Ni_{51}$ alloy as a ferromagnetic material (F). The multi-reentrant state, expected theoretically [20] seems to be present, too, in these experiments [22-24].

The same phenomena have recently been demonstrated in F/S bilayers of the same materials, where the superconducting film now is grown on top of the ferromagnetic one [24], contrary to the former investigated S/F bilayers. During the investigation we realized that the layers thickness and fitting parameters of the F/S bilayers, at which the oscillatory and reentrant behavior of superconductivity were observed, are not equal to that of the S/F bilayers, probably, because of different growth conditions.

Thus, we could realize both building blocks necessary for the fabrication of the core structure of the superconducting spin valve, nominally consisting of two mirror symmetric bilayers [25]. In other words, the spin valve core consists of a F/S/F trilayer, which can be regarded as a package of a F/$\tilde{S}$ and $\tilde{S}$/F bilayers so that S=2$\tilde{S}$ in the trilayer [22,24].

However, the real system is not symmetric with respect to the median plane because of different growth conditions of the F layers and of different properties of the F/$\tilde{S}$ and $\tilde{S}$/F interfaces as well. Moreover, in reality an F/S/F trilayer is not simply a stack of the F/$\tilde{S}$ and $\tilde{S}$/F bilayers, studied separately before. The properties of the superconducting layer with a thickness of $d_S=2d_{\tilde{S}}$ (e.g. the electron mean free path and the critical temperature of a free standing film, which increase with the thickness) differ from those of the $\tilde{S}$ layers. Thus, a different influence especially on the reentrant superconducting behavior is expected.

For such a trilayer, the theory [25] predicts that the critical temperature depends on the relative orientation of the magnetization of the ferromagnetic layers. To enable a reversal of one of the magnetizations of the layers with respect to the other by an external magnetic field, the coercive forces of the F layers have to be different due to either intrinsic properties or to an antiferromagnetic pinning layer delivering an exchange bias. Exchange biasing of one of the F layers is a separate problem to be addressed for a forthcoming study.

Up to now only critical temperature shifts in the millikelvin range could be obtained in experimental realizations of such a spin valve [26-42]. We believe that the non-optimal choice of the layers thickness and/or the layers materials could be a reason for the small size of the spin-valve effect in the studies cited above. As we discussed in detail in Ref. [22], with careful tuning of the layers thickness we expect a shift in the Kelvin range for spin valves applying Nb as a superconductor and $Cu_{41}Ni_{59}$ as magnetic material, especially if deep $T_c$ oscillations or even reentrant superconductivity could be realized in the $Cu_{41}Ni_{59}/Nb/Cu_{41}Ni_{59}$ core structure of the valve as a function of the F-layers thickness.

In the present paper we show experimental results on this type of F/S/F trilayer core structure, demonstrating clear oscillations and reentrant behavior of superconductivity. Fitting of the curves has been performed by the theory for non-symmetric F/S/F trilayers given in the Appendix of the present work.

## 2. Sample Preparation and Characterization

### 2.1 Thin Film Deposition and Sample Configuration

The spin valve core can be regarded as a F/$\tilde{S}$-$\tilde{S}$/F structure, as already discussed in the Introduction. Applying our wedge technique, described in detail in Refs. [21,22,43], we fabricated two types of sample series (Fig. 1). In the first one, (a "single wedge" geometry, see Fig. 1a)) the bottom F-layer (made of $Cu_{41}Ni_{59}$ grown on the Si buffer layer) has a constant thickness as well as the subsequently grown $2\tilde{S}$-layer (made of Nb). The F-layer on top (made of $Cu_{41}Ni_{59}$) has a wedge like shape. The second type (a "double wedge" geometry, see Fig. 1b)) of sample series consists of two $Cu_{41}Ni_{59}$ wedges separated by a Nb layer of constant thickness. We used a (80×7mm$^2$ size) commercial (111) Si substrate on which a sample series was magnetron sputtered at the same run. To get a series of 30-35 separate samples cuts of equal distance (about 2.5 mm) were made, following the long side of the substrate, perpendicular to the wedge.

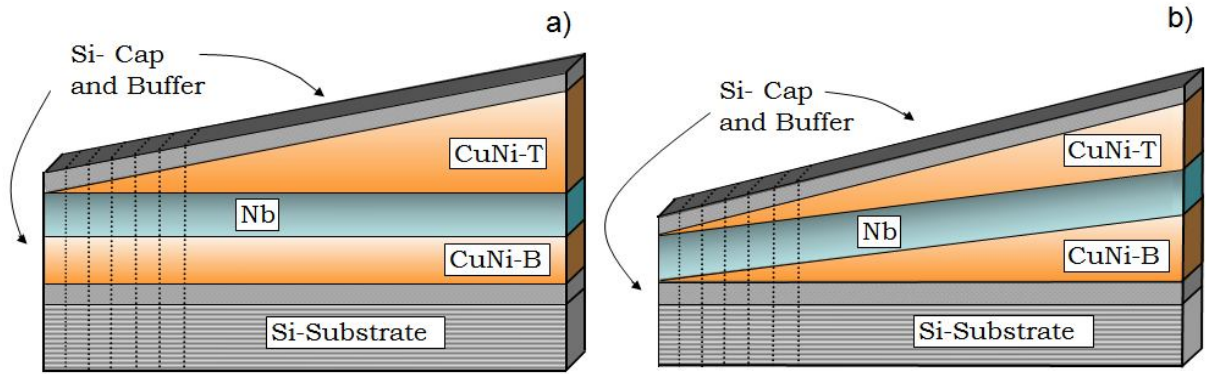

**Fig. 1:** Wedge technique to fabricate $Cu_{41}Ni_{59}/Nb/Cu_{41}Ni_{59}$ trilayers. a) Single wedge geometry: Bottom $Cu_{41}Ni_{59}$ layer ("CuNi-B") kept at a constant thickness. Top $Cu_{41}Ni_{59}$ layer ("CuNi-T") wedge-type. b) Double wedge geometry: Both $Cu_{41}Ni_{59}$ layers are wedges. The symmetric situation is drawn, where the slope of both wedges is the same. In both geometries the thickness of the Nb layer is constant.

## 2.2 Thickness Analysis

Rutherford Backscattering Spectrometry (RBS) was applied to determine the thickness of the individual layers. The $He^{++}$ ions were accelerated to an energy of 3.5 MeV by a tandem accelerator. In order to avoid channeling effects the samples were tilted azimuthally by 7° and the backscattered ions were detected under an angle of 170° with respect to the incident beam.

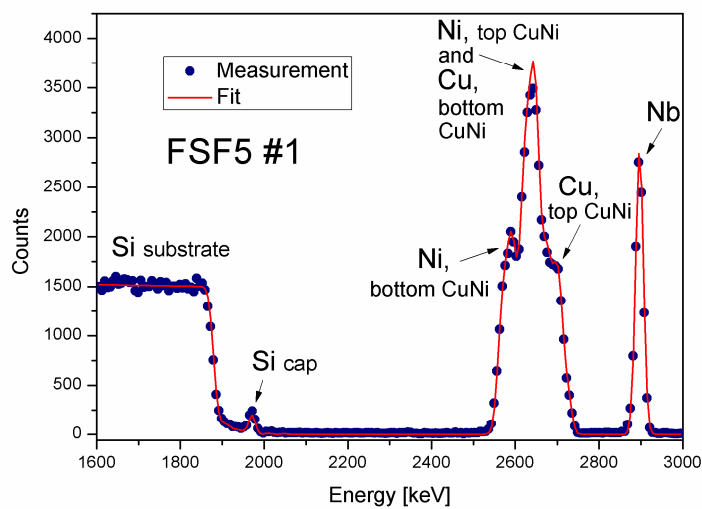

**Fig. 2:** RBS spectrum of sample No1 ("#1") of the FSF5 trilayer series. The blue dots are measurement points whereas the red full line represents the fit. Here, *top* and *bottom* refers to the top and bottom ferromagnetic alloy layer.

The RBS spectrum in Fig. 2 was taken from the double wedge series FSF5. As the Nb layer has a thickness in the nanometer range, the resulting energy shift of the backscattered $He^{++}$ ions is very small, leading to an overlap of the Cu and Ni signal of the bottom and top CuNi alloy layer. To determine the thickness of the top CuNi alloy layer, first the thickness of the silicon cap has to be determined. The energy loss of the ions, which passed through the Si cap layer and which are backscattered at the first CuNi alloy layer, cause the right side of the CuNi alloy peak. The height of the peak results from the combined thickness of the top and bottom CuNi alloy layer, because there is an overlap of the Ni peak of the top alloy layer and the Cu peak of the bottom alloy layer. The left side of this peak is determined by the thickness of the bottom CuNi alloy layer.

From the elemental areal densities obtained by the fitting procedure the thickness of the individual layers can be determined. In contrast to our previous works on bilayers, it is not possible to evaluate a reliable value for the concentration, x, of the $Cu_{1-x}Ni_x$ alloy. However, our experience shows that rf sputtering of the CuNi target used results usually in a x=0.59, i.e. in a $Cu_{41}Ni_{59}$ alloy composition with 41 at. % Cu and 59 at. % Ni (± 1 at. %) of the film, which we, therefore, assume to be present in the CuNi alloy layers for our RBS evaluation.

For the single wedge geometry the RBS spectrum deviates from that shown in Fig 2. In this case the signal of the bottom ferromagnetic layer is almost completely superimposed by the signal of the thicker top layer, i.e. a reliable fitting is hardly possible from the RBS measurements alone. To overcome these difficulties, an alternative possibility (not shown in the present work) is to determine the thickness of the bottom $Cu_{41}Ni_{59}$ alloy layer from a High Resolution Transmission Electron Microscope (HRTEM) measurement and set it constant over the whole wedge.

In Fig. 3a the results of an RBS evaluation are shown for the single wedge geometry (sample series FSF1). The thickness of the layers (CuNi-Top, CuNi-Bottom and Nb) has been plotted as a function of the sample number, i.e. the distance from the thick end of the $Cu_{41}Ni_{59}$ alloy wedge. The bottom CuNi alloy film as well as the Nb film have nearly constant thickness. In Fig. 3b) a similar representation is shown for the double wedge geometry. Again, the Nb film has a constant thickness, whereas the thickness of the CuNi alloy wedges change continuously along the sample series. It is remarkable that for a given sample the thickness of both CuNi alloy films is always nearly equal.

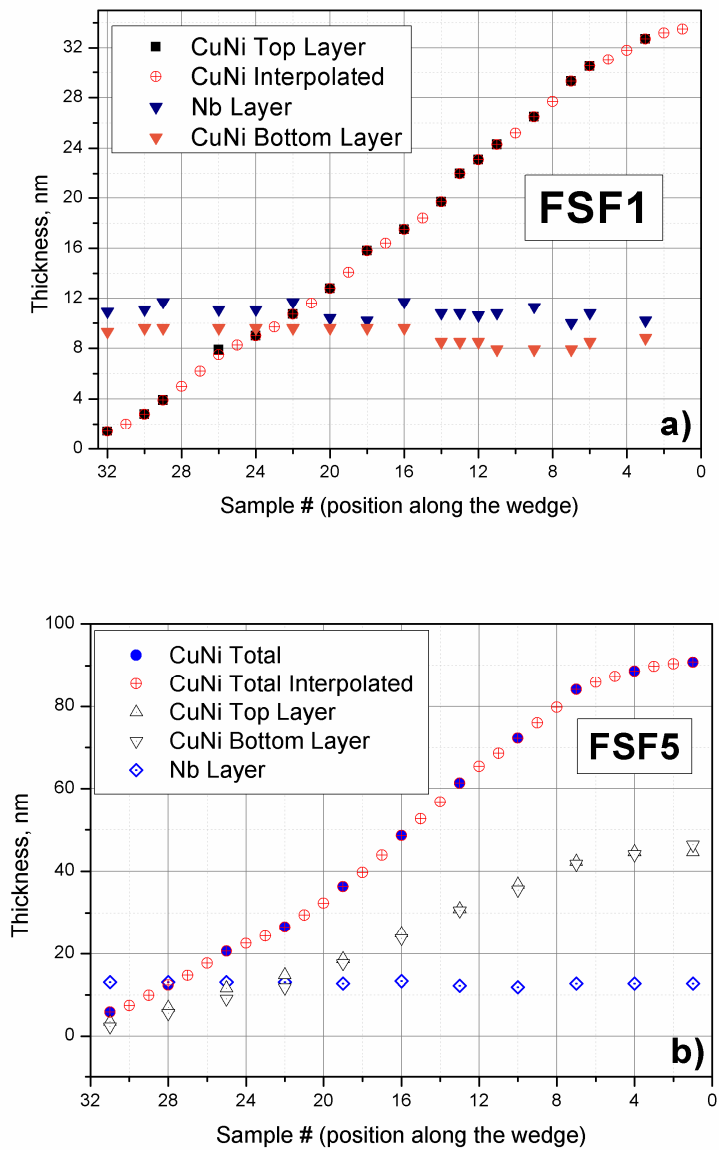

**Fig. 3:** Results of RBS measurements for the thickness of the respective films in F/S/F ($Cu_{41}Ni_{59}$/Nb/$Cu_{41}Ni_{59}$) trilayers plotted against the sample number starting at the thick end of the wedge. a) Single wedge sample series FSF1. Some of the data points of the $Cu_{41}Ni_{59}$ Top layer are interpolated values between two RBS measurements b) Double wedge sample series FSF5. Some of the data points of the sum of the thicknesses of both $Cu_{41}Ni_{59}$ layers ("CuNi Total") are interpolated values between two RBS measurements.

## 2.3 Transmission Electron Microscopy

The growth of our $Cu_{1-x}Ni_x/Nb/Cu_{1-x}Ni_x$ samples has been studied by cross sectional Transmission Electron Microscopy (TEM) using a JEOL JEM 2100F microscope with a GATAN imaging filter and CCD camera. In Fig. 4a) an overview of sample No5 of the Si(substrate)/Si(buffer)/$Cu_{1-x}Ni_x$/Nb/$Cu_{1-x}Ni_x$/Si(cap) single wedge geometry of the FSF1 series is shown. For this sample the interpolated thicknesses (specimen not subjected to the RBS procedure before investigated) from Fig. 3a) are 8.6 nm, 10.6 nm, and 31.2 nm for the bottom $Cu_{41}Ni_{59}$ layer, the Nb layer and the top $Cu_{41}Ni_{59}$ alloy layer, respectively. In Fig. 4b) sample No2 of the double wedge geometry series FSF5 is shown. In this case we have from Fig. 3b) the thicknesses 45.5 nm, 12.7 nm, and 44.6 nm from interpolated RBS data. Both layer thicknesses are in agreement with the TEM picture within about 10%.

The overview over the cross sections of sample FSF1 No5 and FSF5 No2, respectively, show straight and clearly defined boundaries between the different materials over an extended area.

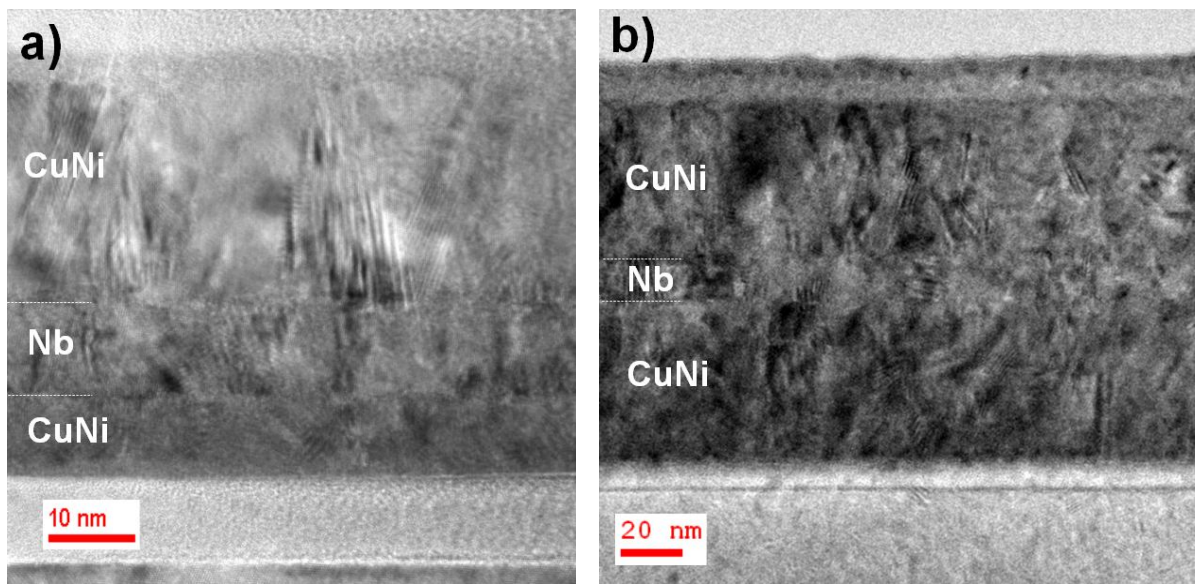

**Fig 4:** Transmission Electron Microscopy images: a) Overview over the cross section of specimen No5, taken from the single-wedge FSF1 sample; b) Overview over the cross section of specimen No2, taken from the double wedge sample FSF5.

In Fig. 5b the diffraction pattern of sample FSF5 No. 2 shows diffraction rings due to the polycrystalline ferro-magnetic Cu41Ni59 alloy layers and the superconducting niobium layer, respectively. The diffraction spots arise from the {111}, {220} and the {200} planes of the single crystalline silicon substrate, resulting in a <110> viewing direction in the cross section [44, 45]. From this result, the perpendicular crystallographic direction of the substrate cannot be definitely concluded. Already from the low indexed directions there are several possibilities, e.g. <110>, <100> and <111>. However, from cross sectional HRTEM of the substrate region we obtain directly that the substrate in the case of sample series FSF5 was prepared with {100} planes parallel to the surface.

Since in the diffraction pattern the 111 spots from silicon are present with a lattice spacing of 3.135 Å, these spots can be used as internal standards. Note that the distance from the central spot to these reflections corresponds to the inverse of the interplanar spacing in real space. By doing so, we can properly index the diffraction rings as belonging to $Cu_{41}Ni_{59}$ or Nb.

The polycrystalline Cu41Ni59 alloy gives rise to diffraction rings which can be attributed to the {111}, {200} and {220} planes. This means that in this cross section of the sample one looks along the <110> direction of the face centered cubic Cu41Ni59 lattice [44–48]. Regarding only low indexed planes densely populated by atoms, this yields a <110>, <100>, or <111> direction (with the planes of the same Miller indices perpendicular to thedirections) in the plane view. However, from Fig. 5a we get an inter-planar spacing for Cu41Ni59 lattice planes parallel to the Nb/Cu41Ni59 boundary, indicating a preferred growth with <111> planes parallel to the Si substrate, as already visible in the electron diffraction pattern. Niobium presents a diffraction ring, which corresponds to {110} lattice planes which, according to the diffraction ring size, are separated by 2.38 Å in real space. This means that the bond length has been increased by 2.1%, which would be a rather high value in bulk material. However, it seems to be possible in thin films [49]. The appearance of the 110 diffraction ring along with the absence of the 200 ring indicates that the viewing direction is along the {111} direction [44, 45], i.e. that (regarding only low indexed planes with a dense package of atoms) the bcc niobium film grows with the {110} planes parallel to the substrate. Indeed {110} planes are visible in Fig. 5a for niobium, according to the inter-planar lattice spacing of 2.38 Å. This, moreover, represents a direct observation of the increased bond length, already concluded from the electron diffraction pattern. The results are in agreement with the growth of Cu41Ni59 and Nb observed by X-ray diffraction for $Nb/Cu_{41}Ni_{59}$ bilayers (not shown here) and the orientation observed by HRTEM for Nb in [24].

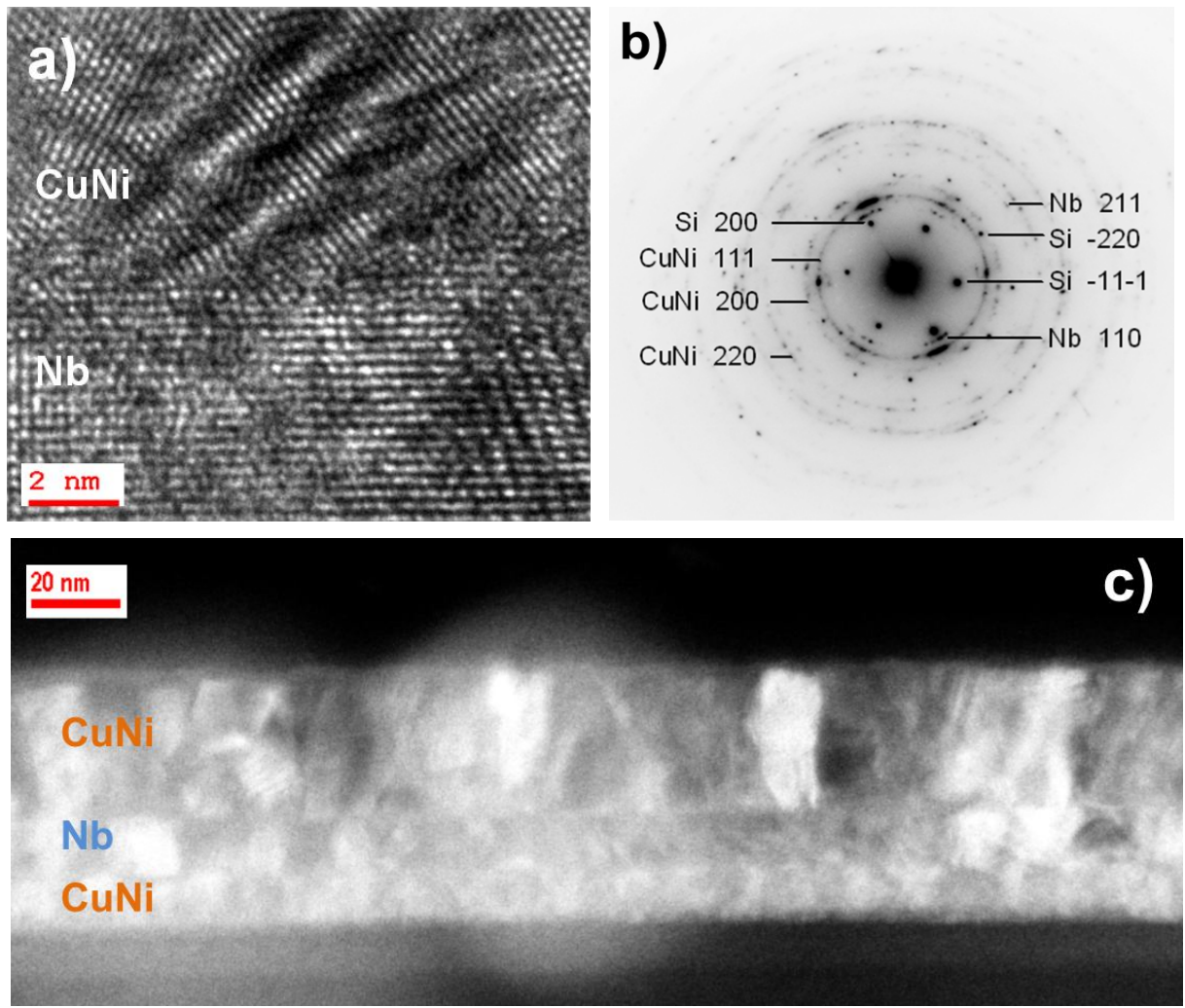

**Fig 5:** Transmission electron microscopy images and electron diffraction pattern: a) High resolution image of the superconductor-ferromagnet interface of sample FSF5 No 2; b) electron diffraction pattern of Sample FSF5 No2; c) Dark field image of the FSF1 No 5 specimen;

From this analysis we get that the $Cu_{41}Ni_{59}$ layers grow with {111} planes parallel to the {100} silicon substrate surface during the deposition, and that the intermediate niobium layer grows with {110} planes parallel to the substrate. Since there is an amorphous silicon buffer layer deposited on the single crystalline substrate, there is no mismatch problem between the substrate and the bottom $Cu_{41}Ni_{59}$ alloy layer. That is different for the growth of the niobium layer on top of the bottom ferromagnetic layer. In this case the growth of the niobium has to occur with an orientation which minimizes the interface energy. The same is the case for the growth of the upper $Cu_{41}Ni_{59}$ layer on top of the Nb layer. For fcc(111)/bcc(110) interfaces several preferred relative orientations of the planes with respect to each other are discussed in the literature [50–53]. These are the Nishiyama-Wassermann (NW) [50–53] orientation where the $[1\bar{1}0]_{fcc}$ or $[0\bar{1}1]_{fcc}$, see [50], direction is parallel to the [001]$_{bcc}$ direction, the Kurdjumov-Sachs (KS) [50–53] relationship with [1¯10]$_{fcc}$ parallel to [1¯11]$_{bcc}$ and a R30°orientation relationship [52, 53] with [2¯1¯1]$_{fcc}$ parallel to [001]$_{bcc}$. What kind of relationship is realized depends on the ratio $d_{bcc}/d_{fcc}$ of the bulk atomic diameters of the bcc and fcc lattice [50].

In a rigid sphere model, for a given lattice constant, the atomic diameter is equal to the nearest neighbor distance. For the bcc and fcc lattice this distance has to be determined from the atomic arrangement in the <111> and <110> directions, resulting in $(\sqrt{3}/2)a_{bcc}$ and $(\sqrt{2}/2)a_{fcc}$, respectively. Here $a_{bcc}$ and $a_{fcc}$ are the lattice constants. The ratio of the bulk atomic diameters is, thus, given by $d_{bcc}/d_{fcc} = (\sqrt{6}/2)a_{bcc}/a_{fcc}$. Using 3.303 Å and 3.539 Å for the lattice constants of bcc Nb and fcc $Cu_{41}Ni_{59}$, respectively, it follows $a_{bcc}/a_{fcc} = 0.93$ and, thus, $d_{bcc}/d_{fcc} = 1.14$. For $d_{bcc}/d_{fcc} = 1.14$ (or $a_{bcc}/a_{fcc} = 0.93$, respectively) the interfacial energy of the NW orientation shows a minimum (Fig. 6 of [50]) and thus, this orientation is expected to be realized for our $Cu_{41}Ni_{59}$ fcc {111}/Nb bcc {110} interfaces (Table I of [51]).

Since both $Cu_{41}Ni_{59}$ layers grow in the same orientation with respect to the superconducting niobium layer, the proximity effect across the boundary, e.g. the transparency [20], are expected to be similar for both interfaces.

## 3. Results and Discussion

### 3.1 Superconducting Properties

The transition temperature of the trilayers was obtained as the midpoint of their resistance-temperature transition curves measured in a Helium-3 cryostat and a dilution refrigerator. The standard dc four probe techniques (10µA measuring current for 0.4-10K and 2µA for 40mK-1K) was applied. To eliminate thermoelectric voltages, alternating the polarity of the current during measurement was done.

In Fig. 6a) the resulting transition temperature for the single wedge series is shown as a function of the thickness of the top F-layer, which changes along the sample series. The average thickness of the constant bottom layer is given in addition. For sample series FSF1 and FSF2 the thickness of the bottom $Cu_{41}Ni_{59}$ alloy film is in average 9.0 nm (see Fig. 3a) and 6.2 nm, respectively. In case of the double wedge samples FSF3 and FSF5 shown in Fig. 6b), the sum of the thicknesses of both F-layers is used. For the FSF5 series the thickness of both layers is nearly the same as seen from the RBS data given in Fig. 3. For the sample series FSF3 the thickness of the top layer increases somewhat more strongly along the wedge compared to the thickness of the bottom layer. For sample #1 and #19 of the upper wedge with $d_{CuNi-T}$=47.4nm and 14.7 nm the difference to the lower wedge with the smaller thickness is 7.9 nm and 2.3 nm, respectively. Between these samples both thicknesses show nearly a straight line behaviour as a function of the sample number.

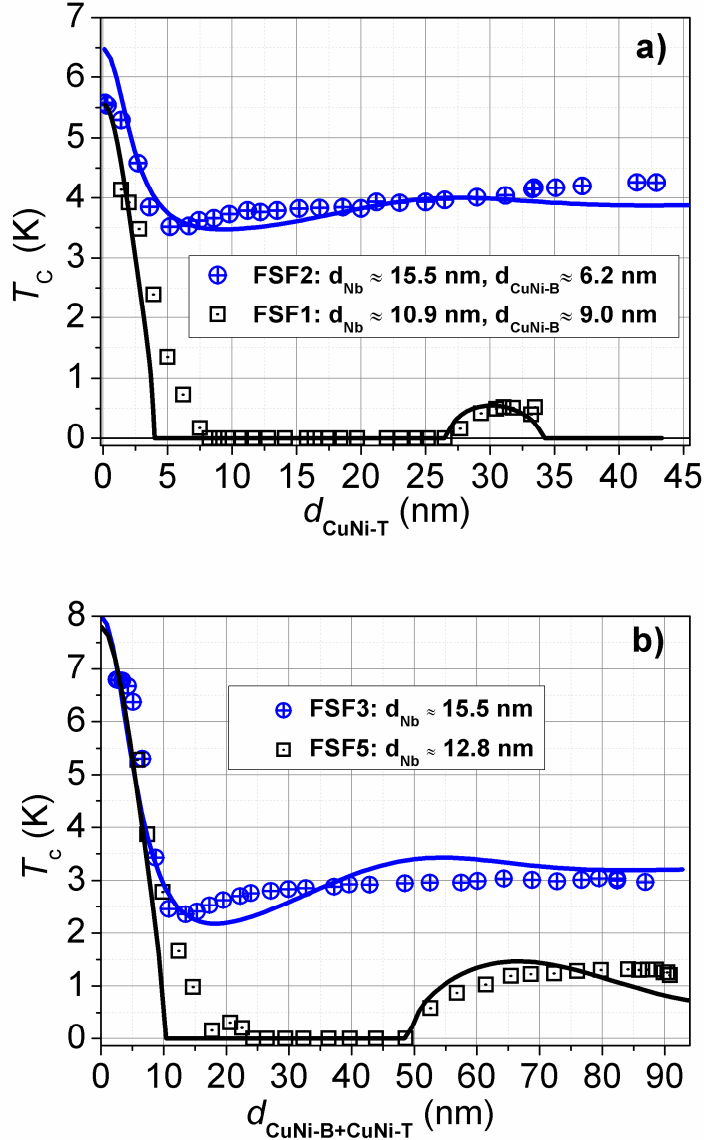

**Fig. 6:** Transition temperatures, $T_c$, of the investigated $Cu_{41}Ni_{59}$/Nb/$Cu_{41}Ni_{59}$ trilayer samples as a function of the thickness $d_{CuNi}$ of the $Cu_{41}Ni_{59}$ alloy layers. a) Single wedge sample series FSF1 and FSF2. Critical temperature as a function of the increasing thickness of the top layer $d_{CuNi-T}$. Thickness $d_{CuNi-B}$ of the bottom flat $Cu_{41}Ni_{59}$ layer and the niobium layer is given in the figure. b) Double wedge series FSF3 and FSF5 plotted as a function of the sum of the thickness of the bottom and top layer, $d_{CuNi-B}$+$d_{CuNi-T}$. The thickness of the niobium layer is given in the figure.

For both types of sample series for a larger thickness of the S-layer an oscillation of the critical temperature is observed. Reducing the thickness of the Nb layer, then yields a reentrant superconducting behaviour.

For samples FSF3 and FSF5 with a double wedge geometry a comparison with results obtained on S/F and F/S bilayers is possible, if we consider that our trilayers are in principle F/S-S/F layers, i.e. if we divide the Nb thickness by a factor of two. Since in average $d_{Nb}$ = 15.5 nm and 12.8 nm for sample FSF3 and FSF5, respectively, we have to compare the behaviour with bilayers having a Nb layer thickness of 7.8 nm and 6.4 nm. For S/F and F/S bilayers indeed an oscillation of $T_c$ is observed for samples with a niobium thickness of about 8 nm, whereas a Nb thickness of about 6 nm leads to reentrant superconducting behaviour [22-24]. Moreover, also the range of $d_{CuNi, AV} = (1/2)(d_{CuNi-B}+d_{CuNi-T})$ in which superconductivity vanishes in the reentrant case of sample series FSF5, is comparable to S/F and F/S bilayers. In the case of sample series FSF3 the minimum of the critical temperature is close to $d_{CuNi, AV}$=6.8 nm, in agreement with S/F bilayers [22].

For the single wedge sample series FSF1 and FSF2 a direct comparison with bilayer measurements is only possible for the CuNi alloy thickness of the upper layer which is equal to the constant lower layer. For FSF1 and FSF2 this thickness is 9 nm and 6.2 nm, respectively, while $d_{Nb}$=10.9 nm and 15.5 nm, yielding 5.5 nm and 7.7 nm, if dividing by two. For the FSF1 series the critical temperature is zero for $d_{CuNi-T}$=9 nm of the top layer, because the sample series shows reentrant behaviour. For the FSF2 series there is a minimum of the critical temperature at $d_{CuNi-T}$ =6.2 nm. This sample series shows an oscillation of the critical temperature. In spite of the thickness of the Nb layers these observations coincide with the behaviour of S/F bilayers [22].

The resistive measurements were performed usually without applying an external magnetic field. To check the influence of the magnetic domain structure on $T_c$, an external magnetic field was applied parallel to the plane of the trilayer to some of the samples at low temperatures. The magnitude of the field was enough to saturate the ferromagnetic layers [22]. Just before the measurements the field was switched off. In this way we adjust a domain structure which is different to samples which did not experience this procedure. No significant shift in $T_c$ was detected which would be visible in Fig. 6.

**3.2 Comparison of Experimental Data with Theory**

The proximity effect at superconductor/ferromagnet (S/F) interfaces differs from that one at a superconductor/normal (S/N) metal interface. The superconducting pairing wave function does not simply decay into the F-metal but oscillates additionally. The reason is that Cooper pairing of electrons in the F-metal occurs with opposite spin, like in a usual s-wave superconductor, but with a non vanishing momentum. The wave length scale of this oscillation is given by $\lambda_{FM}$, i.e. $k_{FM} = 2\pi/\lambda_{FM}$, where $k_{FM}$ is the wave number and $\lambda_{FM}$ is given by the magnetic coherence length $\xi_F$. For a clean ferromagnet, $l_F >> \xi_{F0}$, it is $\lambda_{FM}$ equal to $\lambda_{F0} \equiv 2\pi\xi_{F0} = 2\pi\hbar v_F/E_{ex}$ [55,56], whereas in the dirty case, $l_F << \xi_{F0}$, one has $\lambda_{FM}$ equal to $\lambda_{FD} = 2\pi\xi_{FD} = 2\pi(2\hbar D_F/E_{ex})^{1/2}$ [12,57]. Here, the diffusion coefficient is given by $D_F = l_F v_F/3$ where $l_F$ is the electron mean free path in the ferromagnetic metal, $v_F$ the Fermi velocity in the F-metal, and $E_{ex} << E_F$ is the exchange splitting energy of a free electron like parabolic conduction band, where $E_F$ is the Fermi energy. The decay of the pairing wave function is governed by, $l_F$, in the clean case and, $\xi_{FD}$, in the dirty case [57-59].

The oscillating behavior of the pairing wave function arises from a space dependent phase, generated by the difference of the momenta of the electrons forming the Cooper pairs [12], resulting in interference phenomena at superconductor/ferromagnet interfaces. If the ferromagnetic layer thickness $d_F$ varies, the interference conditions change periodically between constructive and destructive [16], yielding a pairing function flux through an S/F or F/S interface which is periodically modulated as a function of $d_F$. This results in a modulated coupling between S and F layers generating an oscillation of the transition temperature $T_c(d_F)$ [20-24,43].

In trilayers an additional degree of freedom arises from the mutual alignment of magnetic moments in the F layers. In the language of the pairing wave function oscillations the mutual phase of the oscillations depends on the direction of the magnetization in one of the F layers with respect to the other considered as a reference. This gives a channel to control the interference and, thus, to influence the superconducting $T_c$ by varying the magnetic configurations in the F/S/F trilayer. The necessary background for the calculation of $T_c$ of the trilayer with arbitrary thickness and physical properties of the layer has been given in the Appendix to Ref. [22]. A somewhat different form of the equations, used for the calculations in the present paper, can be found in the Appendix to the present work.

Fig. 6a) shows results of calculations for the single wedge sample where the bottom $Cu_{41}Ni_{59}$ layer is flat with $d_{CuNi-B} \approx 9$ nm (sample FSF1) or $d_{CuNi-B} \approx 6.2$ nm (sample FSF2). The top $Cu_{41}Ni_{59}$ layer is a wedge with the thickness varying from 1.4 nm to 33.5 nm and from 1.4 nm to 43 nm, for sample series FSF1 and FSF2, respectively. The calculations are in reasonable agreement with the measurement using a set of physical parameters close to that one obtained from our S/F and F/S bilayers studies [22,24]: (1) FSF1 – $d_{Nb}$=10.9 nm, $T_{c0,Nb}(d_{CuNi} = 0$ nm$) = 7.55$ K; $\xi_S = 6.0$ nm; $N_F v_F/N_S v_S$(bottom) = 0.19 and $N_F v_F/N_S v_S$(top) = 0.22; $T_F$(bottom) = 0.40 and $T_F$(top) = 0.77; $l_F/\xi_{F0}$(bottom) = 0.65 and $l_F/\xi_{F0}$(top) = 1.1; $\xi_{F0}$(bottom) = 10.8 nm and $\xi_{F0}$(top) = 11.4 nm. (2) FSF2 – $d_{Nb}$=15.5 nm, $T_{c0,Nb}(d_{CuNi} = 0$ nm$) = 8.0$ K; $\xi_S = 6.5$ nm; $N_F v_F/N_S v_S$(bottom) = 0.19 and $N_F v_F/N_S v_S$(top) = 0.22; $T_F$(bottom) = 0.45 and $T_F$(top) = 0.61; $l_F/\xi_{F0}$(bottom) = 0.61 and $l_F/\xi_{F0}$(top) = 0.92; $\xi_{F0}$(bottom) = 11.6 nm and $\xi_{F0}$(top) = 10.8 nm. Moreover, $\xi_{BCS} = 42$ nm [61] has been inserted. The meaning of the physical parameters is as follows: $\xi_S$, the superconducting coherence length, see Chap. IIIA of Ref. [22]; $\xi_{F0}$, the coherence length for Cooper pairs in a ferromagnetic metal as introduced above; $l_F$, the mean free path of conduction electrons in the ferromagnetic material; $N_F v_F/N_S v_S$, the ratio of Sharvin conductances at the S/F interface, and $T_F$, the interface transparency parameter. The critical temperature $T_{c0}$ of the Nb layer for a CuNi layer thickness equal to zero is taken from Fig. 5 of Ref. [22], for a free standing niobium film of the given thickness. The critical temperature from which the calculated curve in Fig. 6b starts at $d_{CuNi\text{-}T}$=0 of the top copper-nickel layer is lower than $T_{c0}$. The reason is that the bottom CuNi layer of constant thickness provides an initial suppression of the critical temperature even at $d_{CuNi\text{-}T}$=0 for the top CuNi layer.

Fig. 6b) shows results of the calculation for the double wedge samples. For this calculation it was assumed that the thickness of both ferromagnetic layers is the same, because both $Cu_{41}Ni_{59}$ layers almost synchronously vary in thickness from about 1-2 nm till 45 nm. The agreement with the experimental data is again quite reasonable with a set of the physical parameters close to that obtained from our S/F and F/S bilayers studies [22,24]: (1) FSF3 – $d_{Nb}$=15.5 nm, $T_{c0,Nb}(d_{CuNi} = 0$ nm$) = 8.0$ K; $\xi_S = 7.0$ nm; $N_F v_F/N_S v_S$(bottom) = 0.20 and $N_F v_F/N_S v_S$(top) = 0.22; $T_F$(bottom) = 0.6 and $T_F$(top) = 0.68; $l_F/\xi_{F0}$(bottom) = 0.6 and $l_F/\xi_{F0}$(top) = 0.68; $\xi_{F0}$(bottom) = 11.6 nm and $\xi_{F0}$(top) = 10.8 nm. (2) FSF5 – $d_{Nb}$=12.8 nm, $T_{c0,Nb}(d_{CuNi\text{-}T} = 0$ nm$) = 7.8$ K; $\xi_S = 6.4$ nm; $N_F v_F/N_S v_S$(bottom) = 0.19 and $N_F v_F/N_S v_S$(top) = 0.21; $T_F$(bottom) = 0.68 and $T_F$(top) = 0.83; $l_F/\xi_{F0}$(bottom) = 0.65 and $l_F/\xi_{F0}$(top) = 1.0; $\xi_{F0}$(bottom) = 14.4 nm and $\xi_{F0}$(top) = 12.8 nm.

As it was forecasted in our previous papers [22,24], the physical parameters of the ferromagnetic alloy layers as well as the interfaces between them and the superconducting Nb are different because of different growth conditions. The bottom CuNi layer grows on top of the Si buffer layer, whereas the top CuNi layer grows on top of the Nb layer. The values of the physical parameters obtained from our calculations for trilayers correlate well with the parameters obtained from fitting the S/F and F/S bilayer samples grown in different sequences. The deviations between the theory and the experimental data can be commented as follows: in our calculations we used constant average thicknesses for the flat layers of superconducting Nb and, for the single wedge sample series, also for the bottom CuNi alloy.

There, however, is a small change of the layer thickness (see Fig. 3) of the Nb and bottom CuNi layers along the 80 mm long substrate, which has not been taken into account. Moreover, not all double wedge type sample series have a completely equal thickness along the sample series. If each fitting is done individually without correlation neither to a partner sample series nor to the previous bilayers studies, almost perfect agreement can be obtained for all four sets of measured data. With our modeling we wanted to check the overall consistency of the theory and the experiment, using basically a common set of parameters with a scatter within ±10% around the average value for each. This seems to be possible.

## 4. Conclusion

In the present work the non-monotonous behavior of the critical temperature of $Cu_{41}Ni_{59}$/Nb/$Cu_{41}Ni_{59}$ trilayers is investigated as a function of the thickness of the ferromagnetic material. Oscillations of the critical temperature and also reentrant behavior of the superconducting state could be realized in two different types of sample series. In the first one, the symmetric case, the thickness of the ferromagnetic material of both layers changes equally while the thickness of the superconducting material is held constant. In the second type, the asymmetric case, the thickness of one of the ferromagnetic layers is fixed while the thickness of the second one is varied, again with a constant thickness of the superconducting material.

These experiments represent an important step towards the superconducting spin valve, which needs a Ferromagnet/Superconductor/Ferromagnet core structure with an oscillatory or, more optimal, a reentrant superconducting behavior to realize the theoretically predicted spin switch effect. The exchange biasing of one of the ferromagnetic layers and experimental realization of the spin-valve effect is a subject of our current studies.

**Acknowledgments**

The authors are grateful to S. Heidemeyer, B. Knoblich and W. Reiber for assistance in the TEM sample preparation. The work was supported by the Deutsche Forschungsgemeinschaft (DFG) under the grant No GZ: HO 955/6-1, and in part by the Russian Fund for Basic Research (RFBR) under the grants No 09-02-12260-ofi_m and 11-02-848-a (L.R.T).

**Appendix**

In our recent experiments on S/F and F/S bilayers the experimental data could be described by extending the "dirty" metals theory towards the clean case and apply it to the intermediate range $l_F \approx \xi_{F0}$ [20], representing the crossover region between the dirty and the clean cases, valid for our samples. The detailed justification of the applicability of this extension is given in the Appendix to Ref. [22], and we refer the reader to that paper. Here, for completeness of the presentation we give a closed set of equations with which the $T_c(d_F)$ dependencies presented in Figs. 6a) and 6b) can be obtained.

The reduced critical temperature $t_c = T_c/T_{c0}$ ($t_c \in [0,1]$) in the single-mode approximation is found as a solution of the equation,

$$\ln t_c = \Psi\left(\frac{1}{2}\right) - \operatorname{Re} \Psi\left(\frac{1}{2} + \frac{\phi^2}{2t_c(d_S/\xi_S)^2}\right), \tag{A1}$$

where $\Psi(x)$ is the digamma function, $\phi = k_S d_S$, and $k_S$ is the complex valued propagation momentum of the pairing function in the S layer. Moreover, $d_S$ is the thickness of the superconducting layer. The alignment of magnetizations of the F layers is assumed parallel. Here, $\xi_S = (D_S/2\pi T_{c0})^{1/2}$ is the superconducting coherence length in the S layer, using $\hbar = k_B = 1$, and $T_{c0}$ is the critical temperature of the stand-alone superconducting layer. Moreover, $D_S = l_S v_S / 3$ is the electronic diffusion coefficient in a superconductor with $l_S$ the electron mean free path and $v_S$ the Fermi velocity in the superconducting matieral.

The dimensionless $\phi$ is a solution of the equation

$$[\phi \tan \phi - R_T][R_B \tan \phi + \phi] + [\phi \tan \phi - R_B][R_T \tan \phi + \phi] = 0, \tag{A2}$$

where

$$R_B = \frac{N_{FB} D_{FB}}{N_S D_S} \frac{k_{FB} d_S \tanh(k_{FB} d_{FB})}{1 + \frac{2 D_{FB} k_{FB}}{T_{FB} v_{FB}} \tanh(k_{FB} d_{FB})}, \tag{A3}$$

and

$$R_T = \frac{N_{FT} D_{FT}}{N_S D_S} \frac{k_{FT} d_S \tanh(k_{FT} d_{FT})}{1 + \frac{2 D_{FT} k_{FT}}{T_{FT} v_{FT}} \tanh(k_{FT} d_{FT})}, \tag{A4}$$

the subscript "B" refers to the bottom CuNi layer, whereas the subscript "T" refers to the top CuNi layer. When inserted into equation (A1) the solution of Eq. (A2) for $\phi$ gives the critical temperature $T_c$ of the generally non-symmetric $F_T/S/F_B$ structure for the parallel alignment of the F layers magnetizations. In equations (A3) and (A4) $k_{FB(T)}$ is the complex-valued propagation momentum of the pairing function in the F layers, $d_{FB(T)}$ the thickness of the F-layers, and $N_S$ and $N_{F(B)}$ are the electronic density of states in the S-metal and F-metal, respectively.

The other constituents of equations (A3) and (A4) are defined and parameterized as follows:

$$k_F d_F = \sqrt{\frac{iE_{ex}}{D_F}} d_F = \frac{d_F}{\xi_{F0}} \sqrt{i\frac{\xi_{F0}}{l_F} - 1}, \qquad (A5)$$

where $\xi_{F0} = \hbar v_F / E_{ex}$, and $l_F$ are the coherence length and the electron mean free path obtained as a fitting parameter in Subsection 3.2, $D_F = v_F l_F/(1 + i l_F/\xi_{F0})$.
Next,

$$D_F k_F / v_F = \left(1 - i\frac{\xi_{F0}}{l_F}\right)^{-1/2}, \qquad (A6)$$

and, finally,

$$\frac{d_S}{D_S} = \frac{\pi}{2\gamma} \frac{d_S}{v_S} \frac{\xi_{BCS}}{\xi_S^2}, \qquad (A7)$$

where the following definitions had been used:

$$\xi_S^2 = \frac{D_S}{2\pi T_{C0}}, \qquad \xi_{BCS} = \frac{\gamma v_S}{\pi^2 T_{C0}}, \qquad (A8)$$

$\gamma \approx 1.781$ is the Euler constant. Substitution of equations (A5)-(A8) into equations (A3) or (A4) yields

$$R = \frac{\pi}{2\gamma} \frac{N_F v_F}{N_S v_S} \frac{d_S \xi_{BCS}}{\xi_S^2} \frac{\tanh\left(\frac{d_F}{\xi_{F0}} \sqrt{i\xi_{F0}/l_F - 1}\right)}{\sqrt{1 - i\xi_{F0}/l_F} + (2/T_F)\tanh\left(\frac{d_F}{\xi_{F0}} \sqrt{i\xi_{F0}/l_F - 1}\right)}, \qquad (A9)$$

where the subscript "B" or "T" is implied at every physical parameter which belongs to the bottom or top CuNi layer as well as the bottom or the top interface between the CuNi ferromagnetic and the Nb superconducting layers, respectively. In particular, the set of the physical parameters consists the superconducting coherence length $\xi_S$, which characterizes the inner Nb layer, and four parameters for each of the ferromagnetic layers and their interfaces with the Nb layer: $\xi_{F0i}$, the coherence length for Cooper pairs in a ferromagnetic metal; $l_{Fi}$, the mean free path of conduction electrons in a ferromagnet; $N_{Fi} v_{Fi}/N_S v_S$, the ratio of Sharvin conductance at the S/F interface, and $T_{Fi}$, the interface transparency parameter, where i=B or

T. The initial guesses for the above nine physical parameters have been taken from our bilayers studies in Refs. [22] and [24].

**References**

[1] J. Bardeen, L.N. Cooper, and J.R. Schriefer, Phys. Rev. **106**, 162, (1957).

[2] M. Tinkham, Introduction to Superconductivity (Second Edition, McGraw Hill, NY, 1996).

[3] P. Fulde and R. Ferrell, Phys. Rev. **135**, A550 (1964).

[4] A. I. Larkin and Yu. N. Ovchinnikov, Zh. Eksp. Teor. Fiz. **47**, 1136 (1964) [Sov. Phys. JETP **20**, 762 (1965)].

[5] P. Fulde, Adv. Phys. **22**, 667 (1973), Fig. 22.

[6] A. Bianchi, R. Movshovich, C. Capan, P.G. Pagliuso, and J.L. Sarrao, Phys. Rev. Lett. **91**, 187004 (2003).

[7] R. Lortz, Z. Wang, A. Demuer, P.H.M. Boettger, B. Bergk, G. Zwicknagl, Z. Nakazawa, and J. Wosnitza, Phys. Rev. Lett. **99**, 187002 (2007).

[8] M. Kenzelmann, Th. Straessle, C. Niedermayer, M. Sigrist, B. Padmanabhan, M. Zolliker, A.D. Bianchi, R. Movshovich, E.D. Bauer, J.L. Sarrao, J.D. Thompson, Science **321**, 1652 (2008).

[9] B. Bergk, A. Deumer, I. Sheikin, Y. Wang, J. Wosnitza, Y. Nakazawa, and R. Lortz, Phys. Rev. B **83**, 064506 (2011).

[10] G. Zwicknagl and J. Wosnitza, International Journal of Modern Physics B **24**, 3915 (2010).

[11] G. Zwicknagl and J. Wosnitza, Breaking Translational Invariance by Population Imbalance: The Fulde-Ferrell-Larkin-Ovchinnikov States, in: BCS: 50 years, L.N. Cooper and D. Felmann, Eds. (World Scientific, Singapore, 2011) Chap. 14.

[12] A. I. Buzdin, Rev. Mod. Phys. **77**, 935 (2005).

[13] I. F. Lyuksyutov and V. L. Pokrovsky, Adv. Phys. **54**, 67 (2005).

[14] A. A. Golubov, M. Yu. Kupriyanov, and E. Il'ichev, Rev. Mod. Phys. **76**, 411 (2004).

[15] F. S. Bergeret, A. F. Volkov, and K. B. Efetov, Rev. Mod. Phys. **77**, 1321 (2005).

[16] A. S. Sidorenko, V. I. Zdravkov, J. Kehrle, R. Morari, E. Antropov, G. Obermeier, S. Gsell, M. Schreck, C. Müller, V. V. Ryazanov, S. Horn, R. Tidecks, and L. R. Tagirov in *Nanoscale Phenomena - Fundamentals and Applications.* H. Hahn, A. Sidorenko, I. Tiginyanu, Eds (Springer-Verlag, Berlin-Heidelberg, 2009) Chap. 1.

[17] Hecht E., 2002 , "Optics", 2nd Ed., Addison Wesley, Chapter 9.6.1.

[18] Z. Radović, M. Ledvij, L. Dobrosavljević-Grujić, A.I. Buzdin, and J.R. Clem, Phys. Rev. B, **44** 759 (1991).

[19]  J.S. Jiang, D. Davidovic, D.H. Reich, and C.L. Chien, Phys. Rev. Lett. **74**, 314 (1995).

[20]  L. R. Tagirov, Physica C **307**, 145 (1998).

[21]  V. Zdravkov, A. Sidorenko, G. Obermeier, S. Gsell, M. Schreck, C. Müller, S. Horn, R. Tidecks, and L. R. Tagirov, Phys. Rev. Lett. **97**, 057004 (2006).

[22]  V.I. Zdravkov, J. Kehrle, G. Obermeier, S. Gsell, M. Schreck, C. Müller, H.-A. Krug von Nidda, J. Lindner, J. Moosburger-Will, E. Nold, R. Morari, V.V. Ryazanov, A.S. Sidorenko, S. Horn, R. Tidecks, and L.R. Tagirov, Phys. Rev. B **82**, 054517 (2010).

[23]  A.S. Sidorenko, V.I. Zdravkov, J. Kehrle, R. Morari, G. Obermeier, S. Gsell, M. Schreck, C. Müller, M.Yu. Kupriyanov, V.V. Ryazanov, S. Horn, L. R. Tagirov, and R. Tidecks, Pis'ma v ZhETF **90**, 149 (2009) [JETP Lett. **90**, 139 (2009)].

[24]  V.I. Zdravkov, J. Kehrle, G. Obermeier, A. Ulrich, S.Gsell, M. Schreck, C. Müller, R. Morari, A.S. Sidorenko, L.R. Tagirov, R. Tidecks, and S. Horn, Supercond. Sci. Technol. **24**, 095004 (2011).

[25]  L. R. Tagirov, Phys. Rev. Lett. **83**, 2058 (1999).

[26]  J. Y. Gu, C.-Y. You, J. S. Jiang, J. Pearson, Ya. B. Bazaliy, and S. D. Bader, Phys. Rev. Lett. **89**, 267001 (2002).

[27]  A. Potenza and C. H. Marrows, Phys. Rev. B **71**, 180503 (2005).

[28]  K. Westerholt, D. Sprungmann, H. Zabel, R. Brucas, B. Hjorvarsson, D. A. Tikhonov, and I. A. Garifullin, Phys. Rev. Lett. **95**, 097003 (2005).

[29]  I. C. Moraru, W. P. Pratt, Jr., and N. O. Birge, Phys. Rev. Lett. **96,** 037004 (2006).

[30]  I. C. Moraru, W. P. Pratt, Jr., and N. O. Birge, Phys. Rev. B **74**, 220507 (2006).

[31]  A. Yu. Rusanov, S. Habraken, and J. Aarts, Phys. Rev. B **73**, 060505 (2006).

[32]  J. Aarts and A. Yu. Rusanov, C.R. Physique **7**, 99 (2006).

[33]  R. Steiner and P. Ziemann, Phys. Rev. B **74**, 094504 (2006).

[34]  D. Stamopoulos, E. Manios, and M. Pissas, Phys. Rev. B **75**, 014501 (2007).

[35]  A. Singh, C. Sürgers, and H. v. Löhneysen, Phys. Rev. B **75**, 024513 (2007).

[36]  A. Singh, C. Sürgers, R. Hoffmann, H. v. Löhneysen, T. V. Ashworth, N. Pilet, and H. J. Hug, Appl. Phys. Lett. **91**, 152504 (2007).

[37]  Dong Ho Kim and T. J. Hwang, Physica C **455,** 58 (2007).

[38]  G. Nowak, H. Zabel, K. Westerholt, I. Garifullin, M. Marcellini, A. Liebig, and B. Hjörvarsson, Phys. Rev. B **78**, 134520 (2008).

[39]  Y. Luo and K. Samwer, Europhys. Lett. **91,** 37003 (2010).

[40] P.V. Leksin, N.N. Garif'yanov, I.A. Garifullin, J. Schumann, H. Vinzelberg, V.E. Kataev, R. Klingeler, O.G. Schmidt, and B. Büchner, Appl. Phys. Lett. **97**, 102505 (2010).

[41] J. Zhu, I. N. Krivorotov, K. Halterman, and O. T. Valls, Phys. Rev. Lett. **105**, 207002 (2010).

[42] P. V. Leksin, N. N. Garif'yanov, I. A. Garifullin, J. Schumann, V. E. Kataev, O. G. Schmidt, and B. Büchner, Phys. Rev. Lett. **106**, 067005 (2011).

[43] A. S. Sidorenko, V. I. Zdravkov, A. Prepelitsa, C. Helbig, Y. Luo, S. Gsell, M. Schreck, S. Klimm, S. Horn, L. R. Tagirov, and R. Tidecks, Ann. Phys. (Berlin) **12**, 37 (2003).

[44] D.B. Williams and C.B. Carter, Transmission Electron Microscopy, (2nd Edition, Springer-Verlag, New-York, 2009).

[45] B. Fultz, J. M. Howe, Transmission Electron Microscopy and Diffractometry of Materials (Springer-Verlag, Berlin, Heidelberg, New York, 2001).

[46] W.P. Davey, Phys. Rev. **25**, 753 (1925).

[47] L. Vegard, Z. Phys. **5**, 17 (1921).

[48] L. Vegard, Z. Kristallogr. **67**, 239 (1928).

[49] Q.D. Jiang, Y.L. Xie, W.B. Zhang, H. Gu, Z.Y. Ye, K. Wu, J.L. Zhang, C.Y. Li, and D.L. Yin, J. Phys.: Condens. Matter **2**, 3567 (1990).

[50] Y. Gotoh and I. Arai, Jpn. J. Appl. Phys. **25**, L583 (1986).

[51] Y. Gotoh and M. Uwaha, Jpn. J. Appl. Phys. **26**, L17 (1987).

[52] Y. Gotoh, M. Uwaha, I. Arai, Appl. Surf. Sci. **33/34**, 443 (1988).

[53] Y. Gotoh and H. Fukuda, Surf. Sci. **223**, 315 (1989).

[54] L. Barns, J. Appl. Phys. **39**, 4044 (1968)

[55] E.A. Demler, G.B. Arnold, and M.R. Beasley, Phys. Rev. B **55,** 15174 (1997).

[56] J. Aarts, J. M. E. Geers, E. Brück, A. A. Golubov, and R. Coehoorn, Phys. Rev. B **56**, 2779 (1997).

[57] Z. Radović, L. Dobrosavljević-Grujić, A. I. Buzdin, and J. R. Clem, Phys. Rev. B **38**, 2388 (1988).

[58] I. A. Garifullin, D. A. Tikhonov, N. N. Garif'yanov, L. Lazar, Yu. V. Goryunov, S. Ya. Khlebnikov, L. R. Tagirov, K. Westerholt, and H. Zabel, Phys. Rev. B **66**, 020505(R) (2002).

[59] J. Aarts, J. M. E. Geers, E. Brück, A. A. Golubov, and R. Coehoorn, Phys. Rev. B **56**, 2779 (1997).

[60] L. Lazar, K. Westerholt, H. Zabel, L. R. Tagirov, Yu. V. Goryunov, N. N. Garifyanov, and I. A. Garifullin, Phys. Rev. B **61**, 3711 (2000).

[61] C. Strunk, C. Sürgers, U. Paschen, and H. v. Löhneysen, Phys. Rev. B **49**, 4053 (1994).